\def\pp{\mbox{PG~1159$-$035}}

\def\v4334{\mbox{V4334 Sgr}}  
  
\def\vv{\mbox{VV 47}}  
\documentclass[]{emulateapj}

\begin{document}

\title{On the possible existence of short-period $g$-mode instabilities 
powered by nuclear burning shells in post-AGB H-deficient (PG1159-type) 
stars}

\author{A. H. C\'orsico\altaffilmark{1}}
\author{L. G. Althaus\altaffilmark{1}}
\author{M. M. Miller Bertolami\altaffilmark{1}}
\affil{Facultad de Ciencias Astron\'omicas y Geof\'{\i}sicas,
       Universidad  Nacional de  La Plata,  
       Paseo del  Bosque s/n,
       (1900) La Plata, 
       Argentina}
\email{acorsico@fcaglp.unlp.edu.ar}
\email{althaus@fcaglp.unlp.edu.ar} 
\email{mmiller@fcaglp.unlp.edu.ar} 
\author{J. M. Gonz\'alez P\'erez}
\affil{Instituto de  Astrof\'isica de Canarias, 38200 La
       Laguna, Tenerife, Spain}
\email{jgperez@iac.es}
\and
\author{S. O. Kepler}
\affil{Instituto de F\'\i sica,
       Universidade Federal do Rio Grande do Sul,
       91501-970 Porto Alegre, RS, 
       Brazil}
\email{kepler@if.ufrgs.br}
\altaffiltext{1}{Member of the Carrera del Investigador Cient\'{\i}fico y 
                 Tecnol\'ogico, CONICET (IALP), Argentina}
\begin{abstract}

We  present   a  pulsational   stability  analysis  of   hot  post-AGB
H-deficient pre-white dwarf stars with active He-burning  shells.  The stellar
models   employed    are   state-of-the-art   equilibrium   structures
representative of PG1159 stars  derived from the complete evolution of
the  progenitor stars,  through the  thermally pulsing  AGB  phase and
born-again  episode.  On  the  basis of  fully nonadiabatic  pulsation
computations, we  confirmed theoretical evidence  for the existence
of a separate PG1159 instability strip in the $\log T_{\rm eff} - \log
g$  diagram characterized  by  short-period $g$-modes  excited by  the
$\epsilon$-mechanism.  This instability  strip partially  overlaps the
already  known GW  Vir instability  strip of  intermediate/long period
$g$-modes destabilized  by the classical  $\kappa$-mechanism acting on
the partial ionization of C and/or  O in the envelope of PG1159 stars.
We found  that PG1159 stars  characterized by thick  He-rich envelopes
and located  inside this overlapping region  could  exhibit both short
and intermediate/long periods simultaneously.

As a natural application of  our results, we study the particular 
case of   \vv,   a  pulsating   planetary   nebula  nucleus   (PG1159-type)
particularly  interesting  because  it  has been reported to exhibit 
a  rich  and  complex
pulsation spectrum including a series of unusually short
pulsation periods.  We found that  the long periods exhibited  by \vv\
can be  readily explained  by the classical  $\kappa$-mechanism, while
the  observed   short-period  branch  below  $\approx   300$  s  could
correspond  to modes  triggered by  the He-burning  shell  through the
$\epsilon$-mechanism, although more observational work is needed to 
confirm the reality of these short-period modes. Were  the existence  of
short-period $g$-modes  in this star convincingly  confirmed by future
observations, \vv\  could be the  first known pulsating star  in which
both  the  $\kappa$-mechanism  and  the $\epsilon$-mechanism  of  mode
driving are \emph{simultaneously} operating.
\end{abstract}


\section{Introduction}  
\label{intro}  

At present,  the study  of stellar pulsations  constitutes one  of the
most fundamental pillars on which the building of stellar astrophysics
rests  on. Although  the theory  of stellar  pulsations  was initially
elaborated to  explain the existence of classical variable
stars such as Cepheids and RR Lyrae, in the last few decades the study
of pulsating stars has been  extended to many other different kinds of
stars that were  regarded as constant stars before  
(e.g., Unno et al. 1989;
Gautschy  \& Saio  1995).  Nowadays,  new classes  of  pulsating stars
across the HR diagram  are being routinely uncovered from ground-based
observations as well  as space missions (e.g., CoRoT,  MOST; see Aerts
et al. 2008).  The study of stellar pulsations through the approach of
asteroseismology  constitutes a  powerful tool  to probe  the internal
structure and evolution of stars.

Most of  the pulsations exhibited by pulsating  stars are self-excited
through  the  classical  $\kappa$-mechanism  operating  in  a  partial
ionization  zone  near  the  surface  of  stars  (Cox  1980;  Unno  et
al. 1989).   As a  matter of fact,  this mechanism is  responsible for
pulsations  of the  stars in  the classical  instability strip  due to
partial ionization of H and HeI  and/or HeII. In the driving zone, the
opacity perturbation increases outward so that radiative luminosity is
blocked  in the  compression phase  of pulsation.   The  region gains
thermal energy in the compression phase and it loses thermal energy in
the expansion phase.

A less  common ---and consequently less  explored--- pulsation driving
mechanism in stars is  the $\epsilon$-mechanism. This mechanism is due
to  vibrational instability  induced by  thermonuclear  reactions.  In
this  case, the driving  is due  to the  strong dependence  of nuclear
burning on  temperature.  During  maximum compression, the
temperature and  thus the nuclear  energy production rates  are higher
than at equilibrium.   So, in the layers where  nuclear reactions take
place,  thermal energy  is gained  at compression  while  the opposite
happens during  the expansion  phase (Unno et  al.  1989;  Gautschy \&
Saio 1995).  An excellent historical account of studies on vibrational
destabilization of  stars by the $\epsilon$-mechanism,  can  be found in
Kawaler (1988)  ---we refer  the interested reader  to that  paper for
details.

In this paper, we  explore the $\epsilon$-mechanism in connection with
pulsating PG1159  stars. These stars, also  called GW Vir  or DOV, are
very  hot H-deficient  post-Asymptotic Giant  Branch (AGB)  stars with
surface  layers rich  in He,  C  and O  (Werner \&  Herwig 2006)  that
exhibit multiperiodic luminosity  variations with periods ranging from
300  to 6000  s, attributable  to non-radial  $g$-modes driven  by the
$\kappa$-mechanism acting on the region of partial ionization of C and
O  (Starrfield et  al. 1983,  1984,  1985; Gautschy  1997; Quirion  et
al. 2004;  Gautschy et  al. 2005; C\'orsico  et al.  2006;  Quirion et
al. 2007).  Some pulsating PG1159 stars are still embedded in a nebula
and  are called  Planetary Nebula  Nuclei Variable  (PNNV)  stars (see
Winget  \&  Kepler 2008  and  Fontaine  \&  Brassard 2008  for  recent
reviews).

Evolutionary  models  of PG1159  stars  with  thick He-rich  envelopes
located at the upper left  portion of the HR diagram are characterized
by the  presence of vigorous  He-burning shells. The first  attempt to
study the  effect of the $\epsilon$-mechanism induced  by a He-burning
shell in  H-deficient pre-white  dwarf stars was  the seminal  work by
Kawaler  et al. (1986).   These authors  found some  $g$-modes excited
through  this  mechanism  with periods  in  the  range  70 to  200  s.
Observationally,  however,  no  signature  of  these  short  pulsation
periods was found in the  surveys of planetary nebula nuclei conducted
at that  time (Grauer et  al. 1987; Hine  \& Nather 1987).   Later on,
stability  analysis on  simplified PG1159  models by  Saio  (1996) and
Gautschy  (1997)  also  predicted  unstable $g$-modes  driven  by  the
$\epsilon$-mechanism with periods in the range $110-150$ s.

The interest in the $\epsilon$-mechanism in the context of H-deficient
post-AGB  stars  has  recently   been  renewed  by  the  discovery  of
luminosity   variations  in   the  PNNV   star  \vv\   ($T_{\rm  eff}=
130\,000 \pm  5000$ K, $\log g= 7  \pm 0.5$, C/He= 1.5  and O/He= 0.4;
Werner \& Herwig 2006) by  Gonz\'alez P\'erez et al.  (2006). The most
outstanding feature  of \vv\  is that its  period spectrum  appears to
include a series of  unusually short pulsation periods ($\sim 130-300$
s),  the shortest periods  ever detected  in a  pulsator of  its class.
Gonz\'alez  P\'erez et  al.  (2006)  (see  also Solheim  et al.  2008)
speculate  that  these rapid  oscillations  could  be  excited by  the
$\epsilon$-mechanism.

In this work, we largely extend  the pioneering work by Kawaler et al.
(1986),  Saio   (1996)  and   Gautschy  (1997)  by   performing  fully
nonadiabatic  pulsation   computations  on  realistic   PG1159  models
extracted  from  full evolutionary  sequences  with  a  wide range  of
stellar masses  and effective temperatures.  In  particular, we gather
strong evidence for  the  existence of  a  new short-period  $g$-mode
instability   strip   of   pulsating   PG1159   stars   due   to   the
$\epsilon$-mechanism.   In addition,   we examine the possibility
that  the short-period  $g$-modes of  \vv\  could be  excited by  this
mechanism. The  paper is organized as  follow: in the  next section we
briefly  describe  the  input   physics  of  the  PG1159  evolutionary
sequences analyzed  and the nonadiabatic treatment  of the pulsations.
In  Sect.   \ref{results}  we  describe the  stability  analysis.   In
Sect. \ref{vv47}  we present the application  to the star  \vv, and in
Sect  \ref{pg1159}  we  discuss  the  case  of  the  prototypical  DOV
star \pp\  in the  context of our  theoretical findings.   Finally, in
Sect.   \ref{summary} we  summarize  our main  results  and make  some
concluding remarks.

\section{Evolutionary/pulsational modelling of PG1159 stars}  
\label{evolutionary}  
  
\begin{figure}  
\centering  
\includegraphics[clip,width=250pt]{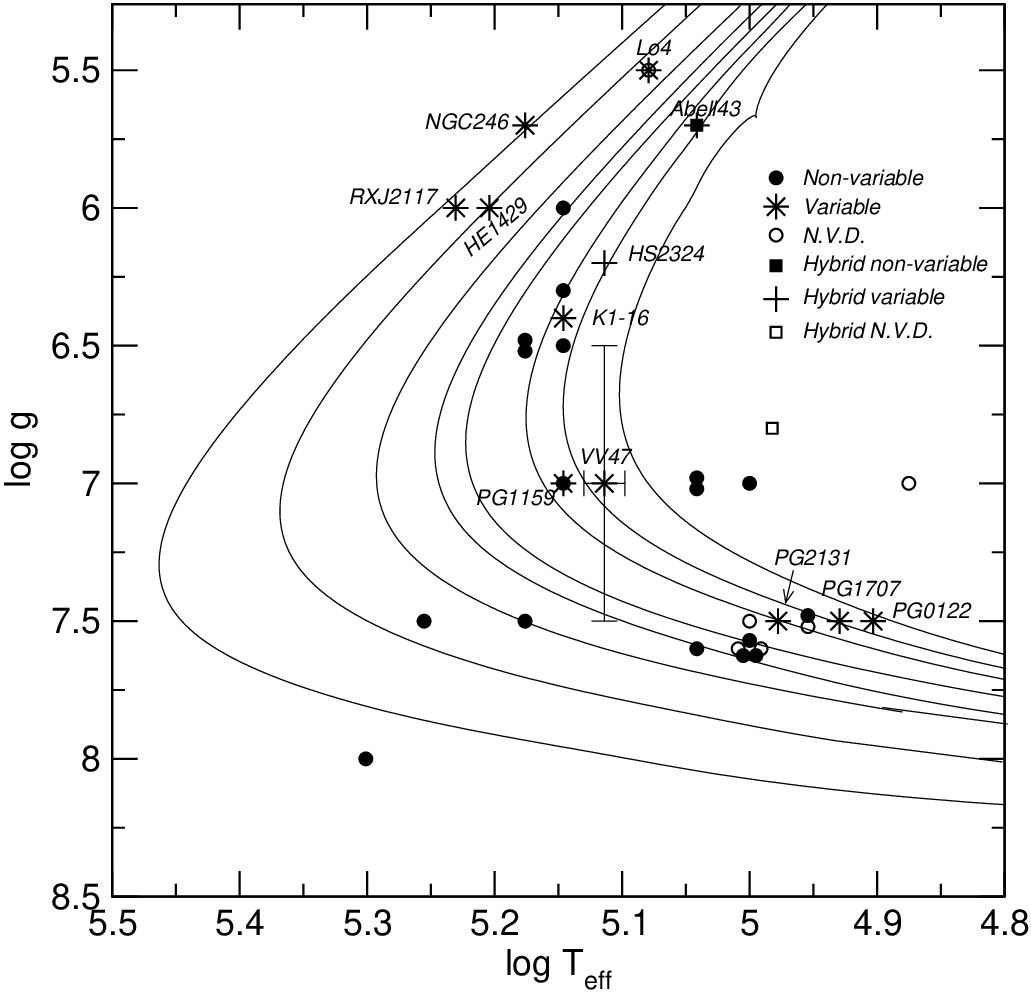}  
\caption{The  PG1159  evolutionary  tracks of Althaus  et  al.  (2005),  
Miller Bertolami \& Althaus (2006)  and C\'orsico et al.  (2006), with
stellar masses  of (from  right to left):  $M_*= 0.515,  0.530, 0.542,
0.565,  0.589,  0.609, 0.664,  0.741  M_{\odot}$.  Also  shown is  the
location  of  known  PG1159  stars.   The  error  bars  for  \vv\  are
displayed.}
\label{fig01}  
\end{figure}

The PG1159 equilibrium models on which the present investigation rests
on  were extracted  from  the evolutionary  calculations presented  by
Althaus  et  al.   (2005),  Miller  Bertolami \&  Althaus  (2006)  and
C\'orsico et al.  (2006), who computed the complete evolution of model
star sequences with initial masses on  the ZAMS in the range $1 - 3.75
M_{\odot}$.   The evolutionary  tracks for  the  H-deficient pre-white
dwarf remnants are displayed in Fig.  \ref{fig01}. All of the post-AGB
evolutionary  sequences, computed  with {\tt  LPCODE} (Althaus  et al.
2005), were  followed through the  very late thermal pulse  (VLTP) and
the resulting  born-again episode that gives rise  to the H-deficient,
He-, C-,  and O-rich composition characteristic of  PG1159 stars.  For
details about  the input physics  and evolutionary code used,  and the
numerical  simulations  performed to  obtain  the PG1159  evolutionary
sequences employed here,  we refer the interested reader  to the works
mentioned above. One distinctive  feature, and crucial for this study,
that is common to all of  our sequences, is that the PG1159 models are
characterized by  He-rich envelopes thick enough as  to sustain active
He-burning   shell   sources  during   the   evolutionary  stages   of
interest.  This is  at variance  with the  non-standard  PG1159 models
employed in  Althaus et al.   (2008) to explain the $\dot{\Pi}$ values
in \pp, which are characterized  by thin
He-rich  envelopes and  so  they are  not  able to  sustain an  active
He-burning shell.

The pulsational stability analysis  presented in this work was carried
out with the linear,  nonradial, nonadiabatic pulsation code described
in   C\'orsico   et  al.    (2006).    The  ``frozen-in   convection''
approximation  was  assumed  because  the  flux  of  heat  carried  by
convection is negligible in  PG1159 stars.  At variance with C\'orsico
et al. (2006), in this work we have fully taken into
account  the $\epsilon$-mechanism  for mode  driving operating  in the
He-shell nuclear-burning region.  Because  we are interested in PG1159
stars which are H-deficient, we are only concerned with the He-burning
reactions.   Fortunately, because  $g$-mode  pulsation timescales  are
much shorter  than the  timescales of nucleosynthesis,  possible phase
delays  between  the   temperature  perturbations  and  the  abundance
variations  are unimportant.   Hence, they  can be  neglected, largely
simplifying  the pulsational  stability analysis  (Unno et  al.  1989;
Kawaler et al. 1986)\footnote{We note  that some of our sequences have
trace  surface abundances of  H ($X_{\rm  H} \lesssim  10^{-3}$) which
give  rise to  some  H burning.   However,  exhaustive test  stability
computations demonstrate that H burning  is very weak and extends on a
extremely narrow  layer, as a result  of which the  H-shell burning is
completely irrelevant  in destabilizing modes and will  not be further
considered in this paper.}.

\begin{figure} 
\centering 
\includegraphics[clip,width=250pt]{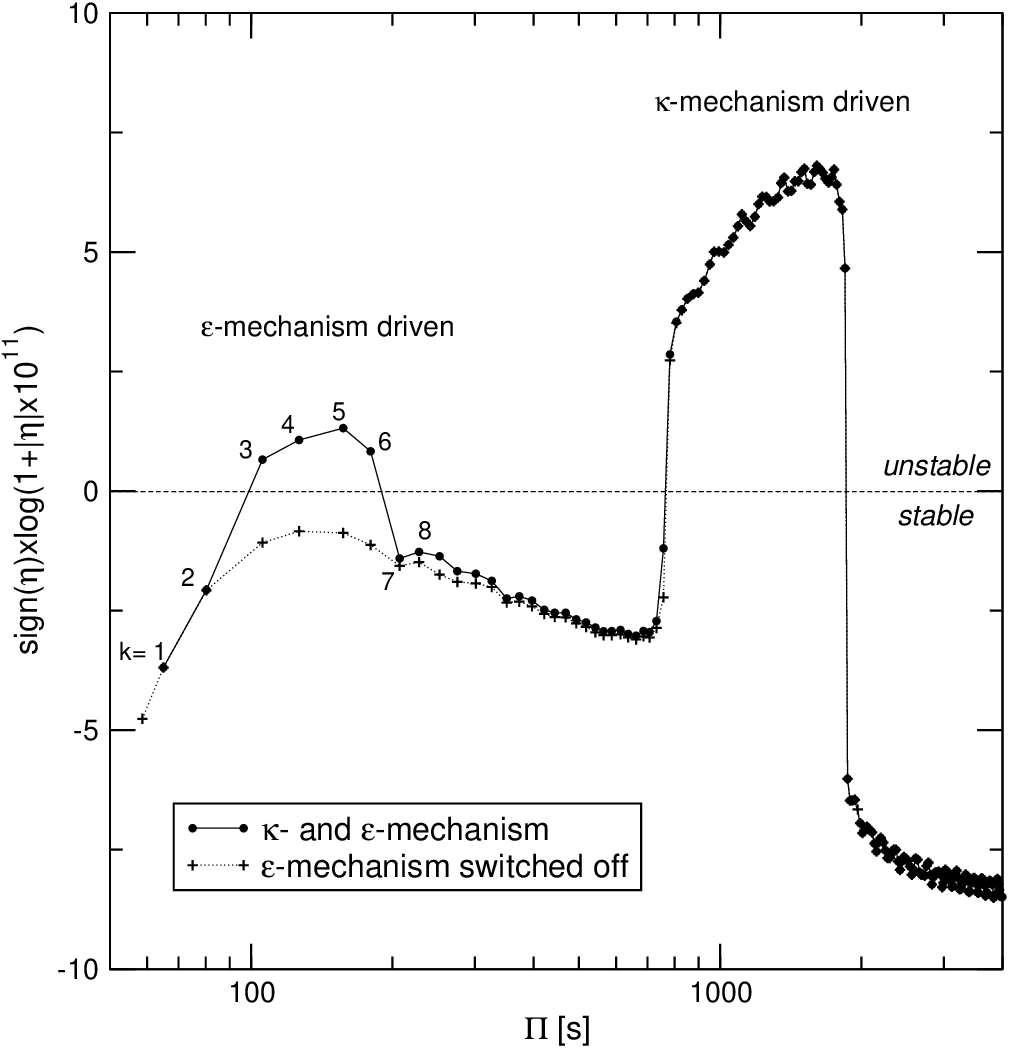} 
\caption{The dipole ($\ell= 1$) normalized growth rates 
$\eta$ (black  dots connected with  continuous lines) in terms  of the
pulsation  periods  for  a  $0.530 M_{\odot}$  PG1159  template  model
located  before the  evolutionary knee  ($T_{\rm eff}=  138\,  400$ K,
$\log (L_*/L_{\odot})=  3.14$). Numbers indicate the  radial order $k$
for low  order modes. The large  numerical range spanned  by $\eta$ is
appropriately scaled  for a better graphical  representation.  The two
ranges of  overstable $g$-modes ---one due  to the $\kappa$-mechanism
and  the  other induced  by  the  $\epsilon$-mechanism--- are  clearly
discernible. Plus  symbols connected  with dotted lines  correspond to
the case  where the  $\epsilon$-mechanism is explicitly  suppressed in
the stability calculations.}
\label{fig02}  
\end{figure} 

\section{Nonadiabatic results}
\label{results}

We  analyzed the  stability properties  of about  4000  stellar models
covering     a     wide     range    of     effective     temperatures
($5.5  \gtrsim  \log(T_{\rm eff})  \gtrsim  4.7$)  and stellar  masses
($0.515  \lesssim M_*/M_{\odot}\lesssim  0.741$).  For  each  model we
restricted our study to $\ell=  1$ $g$-modes with periods in the range
$50-7\,000$ s.
 
\subsection{A single template model}

We start our description by focusing on a $0.530 M_{\odot}$
PG1159  template  model  with  $T_{\rm  eff}= 138\,400$  K  and  $\log
(L_*/L_{\odot})=  3.14$ located  before  the
evolutionary knee in Fig. \ref{fig01}.  The surface 
chemical composition of the model is $X({^4{\rm He}})= 0.33$, 
$X({^{12}{\rm C}})= 0.39$, $X({^{13}{\rm C}})= 0.05$, 
$X({^{14}{\rm N}})= 0.02$, and $X({^{16}{\rm O}})= 0.17$.
Fig.  \ref{fig02}  displays  the
normalized  $\ell= 1$  growth  rates $\eta=  -\Re(\sigma)/\Im(\sigma)$
(where $\Re(\sigma)$ and $\Im(\sigma)$  are the real and the imaginary
parts, respectively, of the  complex eigenfrequency $\sigma$) in terms
of the pulsation periods  ($\Pi$) corresponding to our template model.
In  the  interests of  a  better  graphical
representation,  the  huge  numerical   range  spanned  by  $\eta$  is
appropriately scaled (see Gautschy  1997).  The sign function allows to
discriminate between  stable and unstable modes.  The  presence of two
well-defined   families   of   overstable   $g$-modes,  one   at   the
intermediate- and long-period  regime, and the other one  at the short
period  regime, is  apparent.   The first  group  of periods  ($\approx
750-1800$   s)  corresponds   to  modes   driven  by   the  well-known
$\kappa$-mechanism operating at the region  of the opacity bump due to
partial  ionization  of C  and O, centered  at  $\log  T \approx  6.2$
(Gautschy et  al. 2005, C\'orsico  et al.  2006). The second
group of periods, which are  associated to low radial order $g$-modes,
are  destabilized  by the  action  of  the  vigorous He-shell  burning
through  the  $\epsilon$-mechanism.   

The short-period  instabilities uncovered here are of  the same nature
than those  found  by Kawaler  et al. (1986)  in the  context of
H-deficient hot central stars of  planetary nebulae.  Here, as in that
work,  the  $\epsilon$-mechanism   induced  by  the  He-shell  burning
constitutes the  source of driving.  In absence  of this destabilizing
agent,  the  overstable modes  with  periods  in  the range  ($\approx
100-180$ s)  turn out to be  stable, while the remainder  modes of the
pulsation spectrum  remains unchanged.   This is vividly  displayed in
Fig.  \ref{fig02},  that  shows  with  plus  symbols  the  results  of
additional  stability   computations  in  which   the  nuclear  energy
production   rate,  $\epsilon$,   and   the  logarithmic   derivatives
$\epsilon_T=    \left(\frac{\partial   \ln    \epsilon}{\partial   \ln
T}\right)_{\rho}$                                                   and
$\epsilon_{\rho}= \left(\frac{\partial \ln \epsilon}{\partial
\ln \rho} \right)_T$, are forced to be zero in the pulsation equations.
It is  worth emphasizing  that in  the present effort  we are  able to
obtain    destabilization    of    $g$-modes    through    both    the
$\kappa$-mechanism
\emph{and}  the $\epsilon$-mechanism  in the  same  PG1159 equilibrium
model.  This is at variance with  the study by Kawaler et al.  (1986),
who reported only $\epsilon$-destabilized modes.

 The $\epsilon$-mechanism behaves  as an efficient filter of modes
that  destabilizes  only those  $g$-modes  that  have their  largest
maximum of the temperature perturbation ($\delta T/T$) in the narrow
region of the  He-burning shell (see Kawaler et  al.  1986).  In the
specific case of our template  model, only the $g$-modes with $k= 3,
4, 5$,  and $6$  meet such a  condition and,  as a result,  they are
$\epsilon$-destabilized.   For  $k= 1,  2$  the  largest maximum  of
$\delta T/T$  lies at  inner layers with  respect to  the He-burning
shell.  Thus, these modes are stable.  For modes with $k \geq 7$ the
opposite is true and these modes also are stable.

Test  stability calculations  with $\ell=  2$ for  our  template model
indicate that there  exist only one quadrupole $\epsilon$-destabilized
$g$-mode which  corresponds to $k= 5$  with a period $\Pi  \sim 95$ s,
about $40 \%$  shorter than the corresponding $k=  5$ mode with $\ell=
1$ ($\Pi \sim 157$ s).  Hence, in general, a narrower range of shorter
periods  is  expected to  be  associated with  $\epsilon$-destabilized
$g$-modes with $\ell= 2$ as compared with the case of $\ell= 1$.

\subsection{A new PG1159 instability strip}

\begin{figure}  
\centering  
\includegraphics[clip,width=250pt]{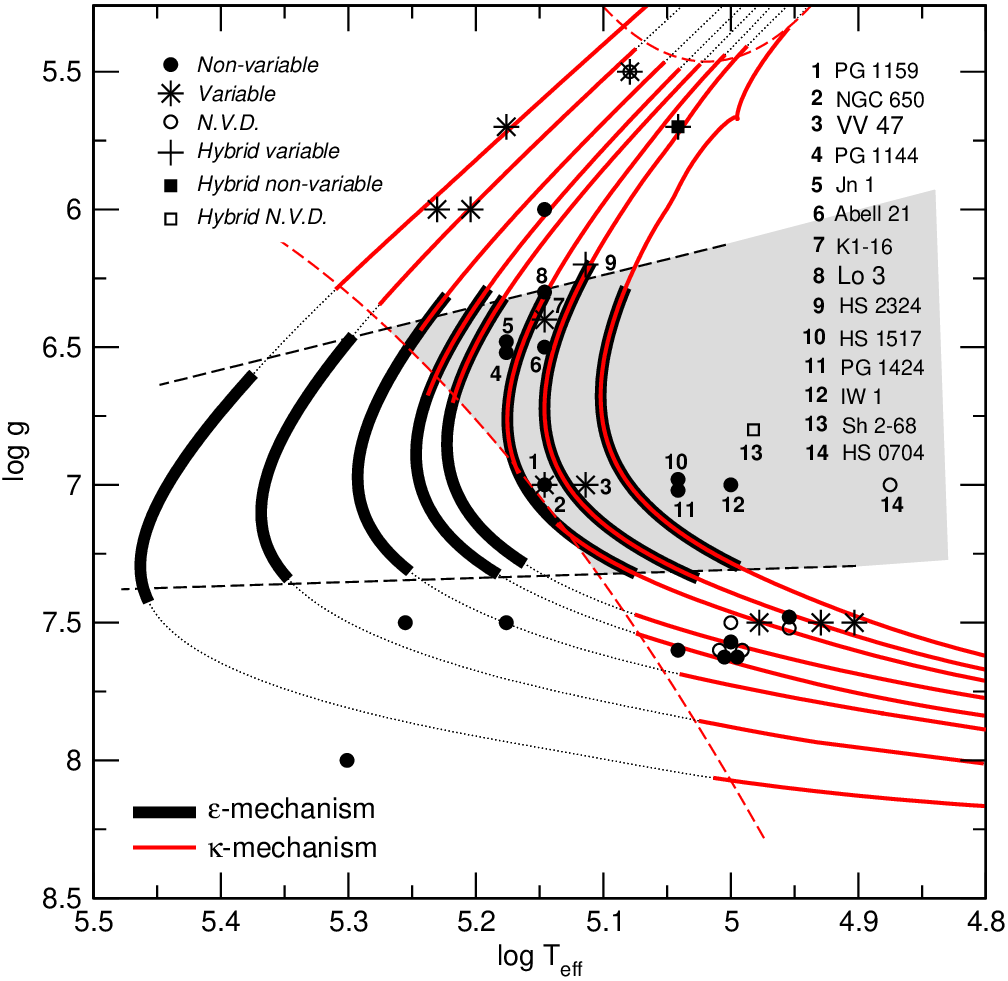}  
\caption{Same as Fig. \ref{fig01}, but including the loci of 
models  having  $\ell= 1$  (dipole)  $\kappa$-destabilized modes  with
solid  red  (gray) curves  along  the  tracks,  and models  harbouring
short-period  $\epsilon$-destabilized modes  according to  the present
study.  Superposition  of both types of curves  corresponds to stellar
models with  both $\epsilon$- and  $\kappa$-destabilized modes (shaded
area).  The location and designation  of relevant PG1159 stars is also
shown [Color  figure only available  in the electronic version  of the
article].}
\label{fig03}  
\end{figure}

Having described our  results for a single template  model, we now are
in conditions  to examine the  location and extension of  the complete
unstable domain associated with the $\epsilon$-mechanism.  Our results
are depicted in Fig. \ref{fig03}, which displays the instability strip
of  $\epsilon$-destabilized modes  in  the $\log  T_{\rm eff}-\log  g$
drawn with  thick black curves  along the PG1159  evolutionary tracks.
In addition,  the GW  Vir instability domain  of $\kappa$-destabilized
modes (see C\'orsico  et al.  2006) is depicted  with red (gray) lines
along   the   tracks.   Note    that   the   instability   strip   for
$\epsilon$-destabilized  modes   partially  overlaps  the   domain  of
$\kappa$-destabilized modes. So, our results indicate the existence of
three well-defined  instability regimes in the  $\log T_{\rm eff}-\log
g$ plane:  a regime ---splitted into  two regions, one  at low gravity
and  the other  at high  gravity---  in which  stellar models  harbour
intermediate/long period  $g$-modes excited by  the $\kappa$-mechanism
only, another one corresponding  to short-period modes destabilized by
the $\epsilon$-mechanism  only, and finally  a region in  which models
experience pulsational  destabilization by the  $\kappa$-mechanism and
the  $\epsilon$-mechanism  of  driving simultaneously.   Notably,  the
region corresponding to the  $\epsilon$-mechanism only is not occupied
by any known PG1159 star (see Fig. \ref{fig03}).

We  stress that in  previous works  (Kawaler et  al. 1986;  Saio 1996;
Gautschy 1997)  only \emph{some} short-period $g$-modes  were found to
be destabilized  by the $\epsilon$-mechanism.  Needles to  say, due to
the very few $\epsilon$-destabilized  modes found in those exploratory
works,  no clear  extension and  location of  the $\epsilon$-mechanism
instability domain  were obtained,  thus hampering those  authors from
making further  consideration of such  modes.  At variance  with those
works, in  the present  study we are  able to find  a \emph{complete}
instability strip of $\epsilon$-destabilized modes.

The degree  of driving, and the place that it  might occur in the
  $\log T_{\rm eff}- \log g$ diagram, is     sensitive to the stellar
  mass, previous  evolutionary history, and  so on.  Thus, due  to the
  uncertainties in  the stellar evolution modelling  (overshooting, nuclear
  reaction rates, etc), the surface and internal composition of PG1159
  stars are not known in detail, and so a clear instability domain for
  $\epsilon$-destabilized pulsations is  difficult to drawn.  So, what
  is shown  in the Fig. \ref{fig03}  is the shape and  location of the
  $\epsilon$-mechanism  instability  strip obtained  by  us under  the
  particular  assumptions adopted  in the  construction of  the PG1159
  evolutionary  models of  Miller  Bertolami \&  Althaus (2006).   The
  extension and  location of this  instability domain might  change if
  other  assumptions for  the evolutionary  history of  the progenitor
  stars were adopted.

All of  the overstable $\epsilon$-destabilized modes  computed in this
work  are  characterized  by  very  tiny  ($10^{-9}-10^{-12}$)  linear
growth  rates  $\eta$,  by   far  smaller  than  those  characterizing
overstable     modes     excited     by     the     $\kappa$-mechanism
($10^{-8} \lesssim  \eta \lesssim  10^{-4}$).  So, the  question rises
about  what would be  the chance  for a  given $\epsilon$-destabilized
mode to have plenty of  time for developing observable amplitudes.  To
analyze this  question we  consider the time  interval that  the models
spend crossing  the instability  strip  of $\epsilon$-destabilized
modes,  $\Delta t$,  and  the maximum  and  minimum $e$-folding  times
$\tau_e^{\rm  max}$  and  $\tau_e^{\rm  min}$,  respectively,  of  the
unstable modes for a given  stellar mass.  The $e$-folding times
are defined as $\tau_e \equiv 1/|\Im(\sigma)|$, such that the time
dependence   of  the  amplitude   of  the   pulsations  is   given  by
$\xi(t)  \propto  e^{i  \sigma   t}$,  and  $\Im(\sigma)  <  0$  for
overstable modes.

The values  of $\Delta t$, $\tau_e^{\rm min}$,  and $\tau_e^{\rm max}$
are provided in Table \ref{table1} for each value of the stellar mass.
Note that  the three timescales monotonically  decrease for increasing
stellar mass. For  all of our PG1159 sequences we  found that the most
unstable  modes ---those with  the smaller  values of  $\tau_e$--- are
found   near  the  low-gravity   (high-luminosity)  boundary   of  the
instability domain (upper black dashed line in Fig. \ref{fig03}), when
the  models  are  still  evolving  to the  blue  before  reaching  the
evolutionary knee.  On the  contrary, when models are already evolving
toward  the  white dwarf  cooling  track, the  $\epsilon$-destabilized
modes are only  marginally unstable, and so they  are characterized by
large $e$-folding times.

Table \ref{table1} shows that $\tau_e^{\rm  min} \ll \Delta t$ for all
of our sequences.  This means  that $g$-modes that are destabilized at
epochs   before  the   evolutionary  knee,   characterized   by  short
$e$-folding times, have time  enough to reach observable amplitudes
before the star  leaves the instability strip.  On  the other hand, it
is  apparent that  $\tau_e^{\rm  max} \gtrsim  \Delta  t$.  Thus,  the
$g$-modes that  are destabilized in  models close to  the high-gravity
limit (low-luminosity)  of the  instability strip (lower  black dashed
line in  Fig. \ref{fig03}) have  little ---or even null---  chances to
develop observable amplitude before the model abandons the instability
domain.

 In  summary, our computations predict that  some $g$-modes (those
  with short $\tau_e$)  could have plenty of time  to grow and finally
  develop  observable  amplitudes.   We  caution, however,  that  this
  prediction is based on  a \emph{linear} stability analysis, and that
  the last word should came  from a detailed non-linear description of
  nonadiabatic  pulsations.    Such  a  nonlinear   treatment  is  not
  available  at the  present  stage.  Also,  there  are other  effects
  (stellar winds, diffusion, etc) suspected to  be present in real stars, that
  could be affecting the growth  of pulsations. The assessment of their
  effects on  the modes predicted to  be unstable in the  frame of our
  analysis is beyond the scope of the present study.

\begin{table}  
\centering  
\caption{The minimum and maximum $e$-folding times (in yr), 
and the time (in yr) that PG1159 models spend within the 
instability strip of $\epsilon$-destabilized modes.}  
\begin{tabular}{cccc}  
\hline  
\hline  
\noalign{\smallskip}
$M_*/M_{\odot}$&$\tau_e^{\rm min}$ & $\tau_e^{\rm max}$& 
$\Delta t$\\  
\noalign{\smallskip}
\hline  
\noalign{\smallskip}
$0.515$ & $3410$ & $1.5 \times 10^6$ & $1.60 \times 10^5$ \\
$0.530$ & $2580$ & $1.0 \times 10^6$ & $1.01 \times 10^5$ \\
$0.542$ & $1610$ & $3.8 \times 10^5$ & $5.95 \times 10^4$ \\
$0.565$ & $1400$ & $1.3 \times 10^5$ & $2.78 \times 10^4$ \\
$0.589$ & $1160$ & $1.0 \times 10^5$ & $2.47 \times 10^4$ \\
$0.609$ & $742 $ & $4.7 \times 10^4$ & $1.26 \times 10^4$ \\
$0.664$ & $361 $ & $1.8 \times 10^4$ & $4830 $            \\ 
$0.741$ & $180 $ & $7000$            & $1570 $            \\ 
\hline  
\end{tabular}  
\label{table1}  
\end{table}  

\begin{figure} 
\centering 
\includegraphics[clip,width=250pt]{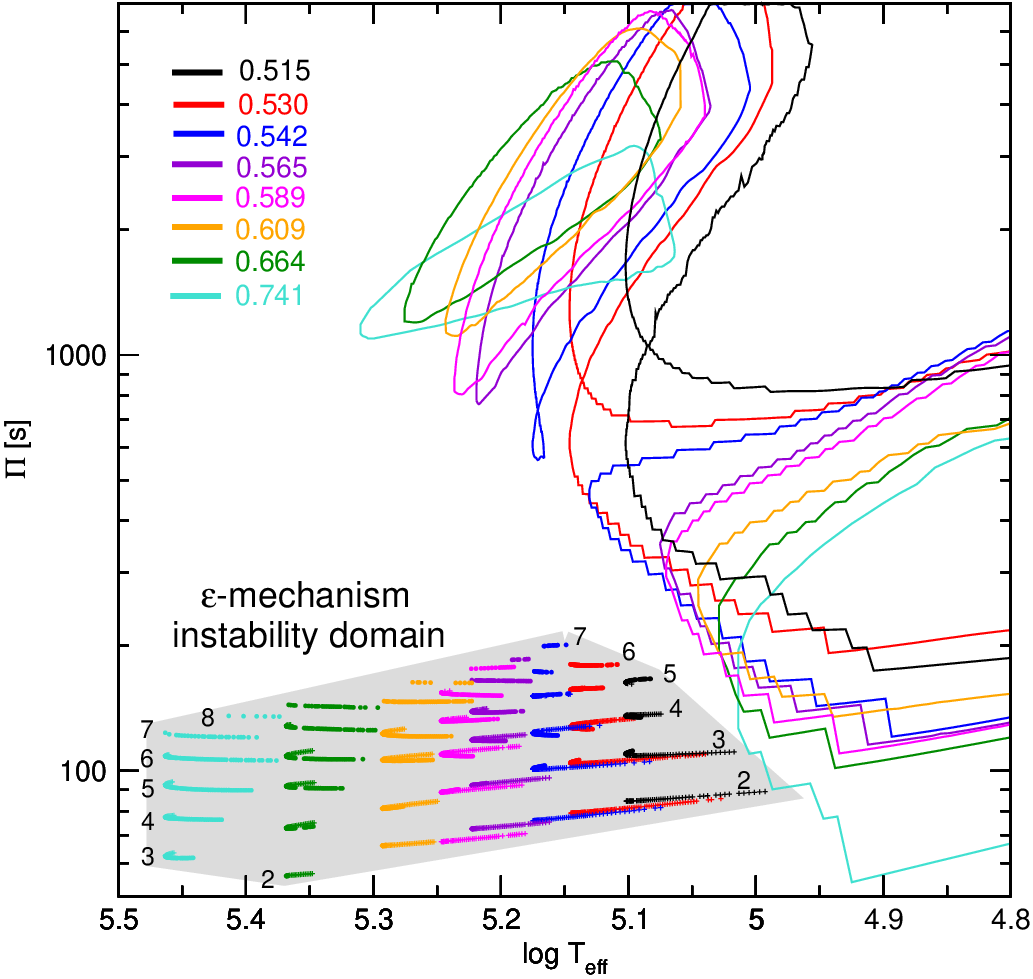} 
\caption{The dipole ($\ell= 1$) instability domains for overstable 
$\kappa$-destabilized  $g$-modes, shown with  thin lines  of different
colours for the various  stellar masses. The $\epsilon$-mechanism 
instability domain is emphasized with a shaded area. 
Short-period dipole unstable
$\epsilon$-destabilized  $g$-modes  are depicted  
with dot  (plus) symbols for stages before (after)  the evolutionary knee.  
Numbers indicate the
radial  order  of the  modes.  [Color  figure  only available  in  the
electronic version of the article].}
\label{fig04}  
\end{figure}

The  next step  in our  analysis  is to  derive the  range of  periods
($\Pi$)      of     overstable      $\epsilon$-destabilized     modes.
Fig.   \ref{fig04}  displays  the  regions of  the  $\kappa$-mechanism
instability domains  in the  $\log T_{\rm eff}-\Pi$  diagram, depicted
with  lines  of different  colours  for  the  various stellar  masses.
Notably,  the   figure  also  shows   the  presence  of   a  separate,
well-defined  instability  domain  for  a  broad  range  of  effective
temperatures  ($5.46 \gtrsim  \log  T_{\rm eff  }  \gtrsim 4.99$)  and
pulsation periods  in the interval  $55 \lesssim \Pi \lesssim  200$ s,
associated  to $\epsilon$-destabilized  $g$-modes  with radial  orders
ranging from $2$ to $5$ for $M_*= 0.515 M_{\odot}$ and from $3$ to $8$
for $M_*= 0.741 M_{\odot}$.  The stages corresponding to phases before
(after)  the evolutionary  knee  are depicted  with  small dot  (plus)
symbols.  A  close inspection  of  the  figure  reveals that  for  the
low-mass models, most of modes are destabilized after the evolutionary
knee. For  the high-mass  models the situation  is reversed,  that is,
most  of overstable  modes  are destabilized  before the  evolutionary
knee.  The  existence of this  new instability domain  of short-period
$g$-modes in stellar models representative of PG1159 stars is the main
result of our study.

In particular,  it is worth emphasizing  that the $\epsilon$-mechanism
should  be  active  in  a  PG1159 star  irrespective  of  the  precise
abundances of  He, C, and O at  the surface, because in  this case the
mode excitation takes place at  deep layers in the star.  This is
at variance  with the $\kappa$-mechanism, which  is strongly dependent
on the exact O/C/He abundances  at the driving regions (see Quirion et
al. 2007).

\subsection{The case of the PNNV star \vv}
\label{vv47}

An immediate prediction of the present study is that PG1159-type stars
populating    the     overlapping    region    of     $\kappa$-    and
$\epsilon$-destabilized modes in the $\log T_{\rm eff}-\log g$ diagram
(the shaded region in Fig. \ref{fig03}) should exhibit both short- and
intermediate/long-period    luminosity    variations   simultaneously.
Table \ref{table2} lists the known PG1159 candidate stars. A glance of
this table  leads us to  a somewhat disappointing conclusion:  most of
the stars located  in the region of interest are  not variables at all
or have not  been scrutinized for variability.  Other  stars, at most,
exhibit  intermediate/long-period   luminosity  variations  which  are
typical  of  the   high/intermediate-order  $g$-modes  driven  by  the
$\kappa$-mechanism,  but not  the  expected short  periods typical  of
$\epsilon$-destabilized modes.   In particular,  this is the  case for
the prototype DOV star, \pp.

There is one object, the PNNV star \vv, which is suspected to
pulsate  in short- and  long-period modes  (Gonz\'alez P\'erez  et al.
2006).  This star ($T_{\rm eff}= 130\,000  \pm 5000$ K, $\log g= 7 \pm
0.5$,  C/He= 1.5  and  O/He= 0.4;  Werner  \& Herwig  2006) was  first
observed as potentially variable by Liebert et al.  (1988).  Later, it
was monitored  by Ciardullo \&  Bond (1996), but no  clear variability
was found.   Finally, Gonz\'alez P\'erez  et al.  (2006) were  able to
confirm the  ---until then,  elusive--- intrinsic variability  of \vv\
for  the first  time.  They  found   evidence that  the pulsation
spectrum  of this  star is  extremely complex.   The  most outstanding
feature of  \vv\ is the  presence of high-frequency peaks  (at periods
$\sim  130-300$ s)  in  the  power spectrum,  which  could be  serious
candidates  for  low-$k$  radial  order  $g$-modes  triggered  by  the
$\epsilon$-mechanism.

We  decided to  test  the attractive
possibility that  the short-periods observed  in \vv\ could be  due to
the  $\epsilon$-mechanism.   We  first  estimated   the  stellar  mass
of \vv.  From the location of \vv\  in the $\log T_{\rm  eff}- \log g$
plane  (see Fig. \ref{fig01})  it is  apparent that  the spectroscopic
mass  of  \vv\  is  of  $\approx 0.525  M_{\odot}$.   In  addition,  a
preliminary  adiabatic  asteroseismological   analysis  on  this  star
indicates  that the seismological  mass of  \vv\ ---obtained  from the
period  spacing data  ($\Delta \Pi  \approx 24$  s)--- is  of $\approx
0.52-0.53  M_{\odot}$, in excellent  agreement with  the spectroscopic
derivation.  So,  we shall  focus  on  the  case of  the  evolutionary
sequence of $M_*= 0.530  M_{\odot}$. This sequence is characterized by
a thick He-rich envelope ($M_{\rm env} \sim 0.045 M_{\odot}$).

\begin{figure} 
\centering 
\includegraphics[clip,width=250pt]{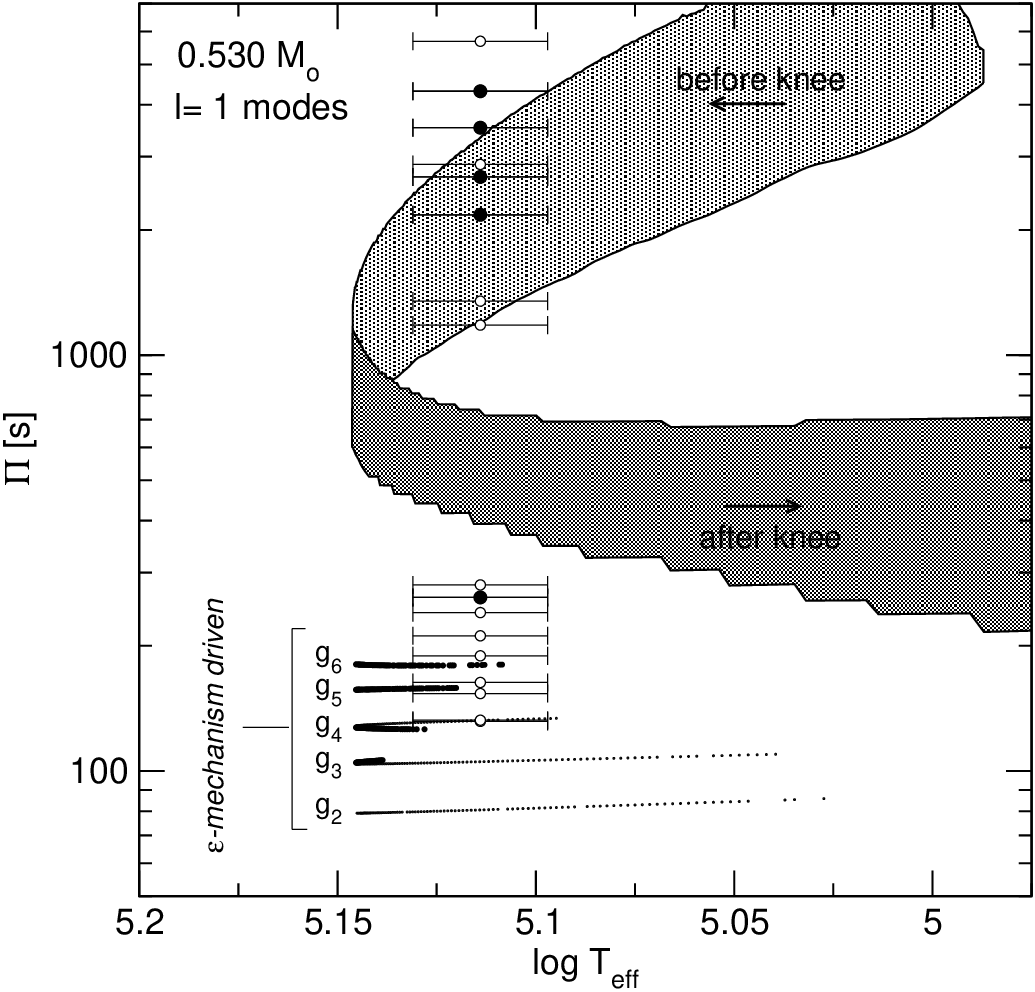} 
\caption{The $\ell= 1$ regions of the
$\kappa$-mechanism  instability  domain,  shown  with light (dark) grey for
stages  before (after)  the  evolutionary knee,  corresponding to  the
$0.530   M_{\odot}$   sequence.    Arrows   indicate  the time sense   of
evolution. Also shown is the evolution of the periods corresponding to
the $\epsilon$-destabilized  modes $g_2, \cdots, g_6$, 
with large (small) dots for stages  before (after)  the  evolutionary knee.
Finally,  the  periods  reported by Gonz\'alez P\'erez et al. (2006) 
for  \vv\  with  their  corresponding
uncertainties in $T_{\rm eff}$ are displayed with small circles.
 Periods detected with sufficiently high significance are emphasized with
black filled circles.}
\label{fig05}   
\end{figure}

We would  like to see  how well the  theoretical ranges of  periods of
unstable modes corresponding to  this sequence fit the observed period
spectrum  of  \vv.  Fig.   \ref{fig05}  displays  the  regions of  the
$\kappa$-mechanism instability  domain (light  and dark grey)  for the
$0.530 M_{\odot}$ sequence.   The figure also shows the  presence of a
well-defined instability domain ($77\lesssim \Pi \lesssim 180$ s) that
corresponds to $\epsilon$-destabilized  $g$-modes with $k= 2,\cdots,6$
(large and  small dots).   Also depicted in  the plot are  the periods
reported  by Gonz\'alez  P\'erez et  al.  (2006)  for \vv\  with their
corresponding uncertainties in $T_{\rm eff}$.   We have emphasized
with black small circles the periods associated with modes having the best
chances to be real, according to Gonz\'alez P\'erez et al.  (2006).
It is apparent that,  
whereas  most of  the  long  periods  observed in  \vv\  are
qualitatively  explained by  the $\kappa$-mechanism  when the model 
star is before the evolutionary knee,  the short-period
branch (below $\sim  300$ s) of the pulsation spectrum  of the star is
not accounted for  at all by the theoretical  domains corresponding to
this destabilizing agent.  We can see, instead, that the short periods
of \vv\  ---in particular $\Pi  \lesssim 200$ s---  are satisfactorily
accounted  for by  the $\epsilon$-destabilized  $g$-modes.  Note,
however,  that  if  only  periods detected  with  sufficiently  high
significance  (black   filled circles)  are   used  to  compare   with  our
theoretical  predictions,  then the  period  at  261.4  s cannot  be
explained    by    a    low-order    $g$-mode   excited    by    the
$\epsilon$-mechanism.  In  fact, this period  is considerably longer
than the longest period ($\approx 180$ s) of the $g$-modes which can
be excited by the  $\epsilon$-mechanism as our analysis predicts. 

Clearly, more observational  work is needed to put  the reality of the
short periods in \vv\ on a solid basis. Were the existence
of  these short periods  confirmed by  future observations,  then they
could  be   attributed  to  the  $\epsilon$-mechanism,   and  this 
could be indicating that \vv\ should have
a  \emph{thick} He-rich envelope  as to  support an  active He-burning
shell. 

\begin{table}  
\centering  
\caption{Known PG1159 stars populating the overlapping 
instability region of $\epsilon$- and $\kappa$-destabilized modes.}  
\begin{tabular}{llccc}  
\hline  
\hline  
&Star             & PN   & Variable  & Period range   [s]\\  
\hline  
1 & PG 1159$-$035 & no   &  yes      & $430-840$   \\
2 & NGC 650$-$1   & yes  &  no       & $-$         \\ 
3 & VV 47         & yes  &  yes (?)  & $\sim$     260  \\ 
  &               &      &           & $\sim$     2170-4300\\ 
4 & PG 1144+005   & no   &  no       & $-$         \\           
5 & Jn 1          & yes  &  yes (?)  & $454-1860$  \\
6 & Abell 21      & yes  &  no       & $-$         \\ 
7 & K 1$-$16      & yes  &  yes      & $1500-1700$ \\  
8 & Longmore 3    & yes  &  no       & $-$         \\        
9 & HS 2324+3944  & no   &  yes      & $2005-2570$ \\
10 & HS 1517+7403 & no   &  no       & $-$         \\
11 & PG 1424+535  & no   &  no       & $-$         \\
12 & IW 1         & yes  &  no       & $-$         \\
13 & Sh 2$-$68    & yes  &  ?        & $-$         \\
14 & HS 0704+6153 & no   &  ?        & $-$         \\
\hline  
\hline  
\end{tabular}  
\label{table2}  
\end{table}

\subsection{The case of the DOV star \pp}
\label{pg1159}

Another consequence  of our investigation concerns  the pulsating star
PG 1159$-$035,  the prototype of  the class and the  best-studied DOV.
Indeed, note  from Fig.   \ref{fig03} that a  trend of our  results is
that    this    variable     star    should    exhibit    short-period
$\epsilon$-destabilized modes if the thick He-rich envelopes derived from our 
evolutionary calculations were representative of the star.
These modes are not observed by Costa et
al. (2007). This  result suggests that the He-burning  shell may not be
active  in \pp. This would  indicate that  this star  has a
thinner  He-rich  envelope than  what is traditionally  derived from  standard
evolutionary calculations, in line with
the recent  finding by  Althaus et al.  (2008) that a  thinner He-rich
envelope (at least a factor of two below of the value predicted by the
standard  evolution theory)  for PG  1159$-$035 should  be  invoked 
to  alleviate the longstanding discrepancy  between the observed
(Costa \&  Kepler 2008) and  the theoretical (C\'orsico et  al.  2008)
rates of period change in that star.

If the short  periods observed in \vv\ were  confirmed, then we should
face  the problem  of  the coexistence  of  two PG1159  stars
located  very close  each  other in  the  $\log T_{\rm  eff} -\log  g$
diagram  (see  Fig.   \ref{fig03})  but with  substantially  different
thickness  of the  He-rich envelopes.  This would  suggest that
these  stars  could  have  had  a different  evolutionary  history,  a
suggestion reinforced by the fact  that \vv\ still retains a planetary
nebula while PG 1159$-$035 does not.

\section{Summary and conclusions}
\label{summary}

In  this  paper, we  have  presented  a  fully nonadiabatic  stability
analysis  on  state-of-the-art  PG1159  models generated  taking  into
account  the  complete  evolution  of progenitor  stars,  through  the
thermally pulsing  AGB phase and born-again episode.  We have explored
the   possibility  that   nonradial  $g$-mode   pulsations   could  be
destabilized by  a He-burning shell  through the $\epsilon$-mechanism.
Our  study  covers a  broad  range  of  stellar masses  and  effective
temperatures. We confirm and  extend the pioneering  work of
Kawaler et al. (1986), Saio  (1996) and Gautschy (1997) on this topic.
The main results are the following:

\begin{itemize}

\item[-] We 
found  strong theoretical evidence  for the  existence of  a separate,
well-defined PG1159 instability strip in  the $\log T_{\rm eff} - \log
g$  diagram characterized  by  short-period $g$-modes  excited by  the
$\epsilon$-mechanism due to the  presence of active He-burning shells.
Notably, this  instability strip partially overlaps  the already known
GW Vir instability  strip due to the $\kappa$-mechanism  acting on the
partial ionization of C and/or O  in the envelope of the PG1159 stars.
We  emphasize that  while  in previous  works  only some  short-period
$g$-modes were  found to be destabilized  by the $\epsilon$-mechanism,
in  the  present  study  we   found  a  \emph{complete}
instability strip of $\epsilon$-destabilized modes.

\item[-] At variance with the classical 
$\kappa$-mechanism responsible for the intermediate/long-period GW Vir
pulsations,  the  $\epsilon$-mechanism  should  be efficient  even  in
PG1159 stars with low C and O content in their envelopes.

\item[-] The $\epsilon$-driven $g$-modes that are destabilized at
epochs  before  the  evolutionary  knee  are  characterized  by  short
$e$-folding times (between $\approx 180$ yr for $M_*= 0.741 M_{\odot}$
and  $\approx  3000$ yr  for  $M_*=  0.515  M_{\odot}$), and  so  they
probably have time enough as to reach observable amplitudes before the
star leaves the instability strip. Note, however, that nonlinear 
effects, or the presence of a variety of phenomena such as stellar winds or 
diffusion, could affect the growth of pulsations.

\item[-] We have closely examined the case of \vv, the only PG1159 star 
for  which  observational evidence  of  the  presence of  short-period
$g$-modes exists  (Gonz\'alez P\'erez et  al. 2006). For this  star we
have  derived for  the first  time  a seismological  mass of  $\approx
0.52-0.53  M_{\odot}$, in  excellent agreement  with  the spectroscopic
mass ($\approx 0.525 M_{\odot}$). If we accept that all of 
the periods reported by Gonz\'alez P\'erez et al. (2006)
are real, our stability analysis provides very
strong  support to  the idea  that the  physical origin  of  the short
periodicities could  be the $\epsilon$-mechanism powered by an
active He-burning shell, whereas  the long-period branch of the period
spectrum  of  this  star  should  be  due  to  the  $\kappa$-mechanism 
acting on the region of partial ionization of C and
O. However, if  only  periods detected  with  sufficiently  high
significance  are   taken into account,  then the  period  at  
261.4  s can not  be explained    by    a    low-order    $g$-mode   
excited    by    the $\epsilon$-mechanism.

\item[-]  We speculate that the absence of short periods 
($\lesssim 300$ s) in the pulsation 
spectrum of PG 1159$-$035 could be indicating that the He-burning shell 
may not be active in this star. This would indicate that 
PG 1159$-$035 has a thinner He-rich envelope than what is 
traditionally derived from standard evolutionary computations.

\end{itemize}

In light of  our results, if the reality of the  short periods of \vv\
were confirmed by follow-up observations, this star could be the first
known   pulsating   PG1159   star   undergoing   nonradial   $g$-modes
destabilized by  the $\epsilon$-mechanism.   Even more, \vv\  could be
the first  known pulsating star  in which both  the $\kappa$-mechanism
and the $\epsilon$-mechanism of mode driving are \emph{simultaneously}
operating. Further  time-series photometry of  \vv\ will be  needed to
firmly establish  the reality of the  short-period pulsations detected
in this star.

On the other hand, the  apparent absence of short-period pulsations in
the remainder variable stars ---such as K 1$-$16, HS 2324+3944, and Jn
1--- could be an indication  that, like \pp, they are characterized by
thin  He-rich envelopes,  as a  result of  which they  should  lack of
stable  He-shell burning.   Another possibility  is  that short-period
pulsations could be  indeed present in these stars,  but with very low
amplitudes, below the actual detection limits.

Also,  it  is  quite  intriguing   the  absence  of  both  short-  and
intermediate/long-period pulsations  in the  up to now  constant stars
(NGC  650$-$1, PG  1144+005, Abell  21, Longmore  3, HS  1517+7403, PG
1424+535,  IW   1)  that  populate  the  overlapping   region  of  the
$\epsilon$- and  $\kappa$-destabilized modes.  In  any case, extensive
searches  for low amplitude intrinsic  variability in  these stars  
and also  in the stars Sh  2$-$68 and  HS 0704+6153, which  have not 
been  observed for
variability yet, should be worth  doing in order to test the existence
of the new
theoretical instability strip uncovered in this work.

  
\acknowledgments

This paper has been benefited from the valuable suggestions and comments 
of an  anonymous referee.   Part of  this work  was supported  by AGENCIA
through the Programa  de Modernizaci\'on Tecnol\'ogica BID 1728/OC-AR,
and by the PIP 6521 grant  from CONICET. This research has made use of
NASA's Astrophysics Data System. Finally, we thank H. Viturro and 
R. Mart\'inez for technical support.

\end{document}